\documentclass[aip,amsfonts,amsmath,amssymb,reprint]{revtex4-2}
\usepackage{graphicx}
\usepackage{dcolumn}
\usepackage{bm}

\usepackage{xcolor}
\usepackage[english]{babel}
\usepackage[utf8]{inputenc}
\usepackage[T1]{fontenc}
\usepackage{mathptmx}
\usepackage{tabularx,booktabs}
\usepackage{soul}
\usepackage{color}

\usepackage{soul, xcolor}
\setstcolor{red}
\begin{document}
	\title{Abrupt transition of the efficient vaccination strategy in a population with heterogeneous fatality rates}

	\author{Bukyoung Jhun}
    \email{jhunbk@snu.ac.kr}
	\affiliation{CCSS, CTP and Department of Physics and Astronomy, Seoul National University, Seoul 08826, Korea}

	\author{Hoyun Choi}
	\affiliation{CCSS, CTP and Department of Physics and Astronomy, Seoul National University, Seoul 08826, Korea}

	\begin{abstract}
	    An insufficient supply of effective SARS-CoV-2 vaccine in most countries demands an effective vaccination strategy to minimize the damage caused by the disease.
	    Currently, many countries vaccinate their population in descending order of age (i.e. descending order of fatality rate) to minimize the deaths caused by the disease; however, the effectiveness of this strategy needs to be quantitatively assessed.
	    We employ the susceptible-infected-recovered-dead (SIRD) model to investigate various vaccination strategies.
	    We constructed a metapopulation model with heterogeneous contact and fatality rates and investigated the effectiveness of vaccination strategies to reduce epidemic mortality.
	    We found that the fatality-based strategy, which is currently employed in many countries, is more effective when the contagion rate is high and vaccine supply is low, but the contact-based method outperforms the fatality-based strategy when there is a sufficiently high supply of the vaccine.
	    We identified a discontinuous transition of the optimal vaccination strategy and path-dependency analogous to hysteresis.
	    This transition and path-dependency imply that combining the fatality-based and contact-based strategies is ineffective in reducing the number of deaths.
        Furthermore, we demonstrate that such phenomena occur in real-world epidemic diseases, such as tuberculosis and COVID-19.
	    We also show that the conclusions of this research are valid even when the complex epidemic stages, efficacy of the vaccine, and reinfection are considered.
	\end{abstract}

	\maketitle
	
	\begin{quotation}
        The effectiveness of vaccination depends highly on the choice of the individuals to vaccinate, even if the same number of individuals are chosen.
        Therefore, effective vaccination strategies have been a central topic of research in mathematical epidemiology and provided quantitative analysis to inform policy-making in the public health domain.
        In this study, we investigated the effectiveness of vaccination strategies to reduce epidemic mortality in a population where the fatality rate varies among groups of individuals by constructing a metapopulation model with heterogeneous contact and fatality rates.
        We show that the effectiveness of vaccination strategies is closely related to the amount of vaccine available.
        When the vaccine supply is low, vaccinating individuals in the descending order of fatality rates is effective; while when the vaccine supply is sufficiently high, vaccinating individuals with high contact rates outperforms the fatality-based strategy.
        By employing simulated annealing, we identify an abrupt transition of the optimal strategy and path-dependency analogous to hysteresis in statistical physics.
        This transition suggests that combining the fatality- and contact-based strategies is less effective than either strategy; therefore, if a country has been employing a specific vaccination strategy, it is inadvisable to convert from that strategy to the other.
        We also show that such phenomena occur in the vaccination of real-world epidemic diseases, such as tuberculosis and COVID-19.
    \end{quotation}

	\section{Introduction}

    The spreading process in complex systems, such as networks~\cite{Dorogovtsev2008,Zimmermann2004,Dorogovtsev2002,Klemm2003,Nekovee2007} and metapopulation~\cite{Watts2005,Colizza2007,Colizza2008,Lloyd2004,Masuda2010}, has been an active field of research for modeling many physical and social phenomena~\cite{Hoang2003,Newman2003,Boccaletti2006}.
    This research has included opinion formation in social groups~\cite{Boccaletti2006,Acemoglu2013,Grabowski2006,Watts2007}, the spread of epidemic diseases~\cite{moreno2002,Ji2014,Watts2005,Pastor-Satorras2001d,Mata2013,DeArruda2020}, and the diffusion of innovations~\cite{Katona2011,Rogers2004,Jhun2019,Jhun2021}.
    Current access to a plethora of data~\cite{Jia2019,Leskovec2007,Fowler2006} on human mobility, collaboration, the contagion of epidemic disease, and temporal contacts, all of which were previously unavailable to researchers, now enable effective research into various dynamic processes in social systems.
    Extensive research devoted to the spreading processes has provided quantitative analyses for policy-making, especially in the public health domain.
    Moreover, study of the spreading process has provided a deeper understanding of phase transitions and critical behaviors, such as the effect of structural heterogeneity on epidemic thresholds~\cite{Watts2005,Pastor-Satorras2001d,Pastor-Satorras2001} and the hybrid phase transition induced by cascades~\cite{Choi2017,Lee2017,Lee2016b}.

    One of the most important topics in mathematical epidemiology is vaccination strategy, which has been extensively studied with various epidemic models~\cite{Pastor-Satorras2002,Hebert-Dufresne2013,Cohen2003,Ghalmane2019,Osat2017,Schneider2011,Costa2020,Yan2015,Clusella2016,Masuda2009,VanMieghem2011,Matamalas2018a,Dong2020,Chen2008,Madar2004}.
    If an individual is vaccinated for certain epidemic disease, that individual acquires immunity to the disease.
    Actual vaccines have less than perfect efficacy, which means that there is a small probability that a vaccinated individual can be infected by the disease (i.e., a vaccine breakthrough).
    It is often modeled that vaccinated individuals do not turn into the infected state even in contact with infected individuals.
    In such a model, when a sufficient fraction of individuals in a system are vaccinated, the infection is unable to spread throughout the system, and the epidemic state is eliminated by the vaccination.
    This effect is called \textit{herd immunity}.
    Vaccination strategies frequently aim to achieve herd immunity with the smallest number of vaccine shots.
    
    The SARS-CoV-2 pandemic is ongoing worldwide and has caused more than five million deaths to date.
    Due to the development of effective vaccines for the disease, the epidemic damage of the disease can be greatly reduced.
    However, in most countries, especially developing countries, the number of vaccine shots available is less than the total population~\cite{Tatar2021}.
    Therefore, it is important to formulate a vaccination strategy that minimizes the damage caused by the disease, such as the number of deaths, with the limited supply of vaccines available.
    Currently, many countries are vaccinating their populations in descending order of age, since the infection fatality rate (IFR) for the COVID-19 increases with age~\cite{Li2020,Levin2020,Manuel2020,Barone-Adesi2020,Kim2020,Shim2021,Bhatt2021}.
    However, the effectiveness of this strategy needs to be quantitatively assessed.

    Here, we employ the susceptible-infected-recovered-dead (SIRD) model, which is a minimal model to study epidemic mortality.
    We evaluate the effectiveness of fatality-based and contact-based vaccination strategies in a metapopulation model with heterogeneous contact and fatality rates.
    We find that the fatality-based strategy is more effective than the contact-based strategy for a high contagion rate and low vaccination supply, but the contact-based strategy outperforms the fatality-based strategy when a sufficiently large amount of vaccine is available.
    Simulated annealing is implemented to find the globally optimal vaccination strategy.
    We find that there is a discontinuous transition of the optimal strategy and path-dependency analogous to hysteresis.
    Further, we demonstrate that these phenomena occur in the vaccination of real-world epidemic diseases, such as tuberculosis (TB) and COVID-19.
    
    This paper is organized as follows: First, we introduce the SIRD model in Sec.~\ref{sec:sird_model}.
    Next, in Sec.~\ref{sec:strategies}, we introduce the synthetic metapopulation model constructed for this research, which has heterogeneous fatality and contact rates.
    In Sec.~\ref{sec:phase_transition}, we study the transition and path-dependency of the vaccination strategy.
    In Sec.~\ref{sec:real_world}, we show that such a transition occurs in real-world epidemic diseases, such as tuberculosis and COVID-19, and in Sec.~\ref{sec:complex_stage}, we demonstrate that the results of this research are valid even when complex details of the epidemic diseases are considered.
    A summary and final remarks are presented in Sec.~\ref{sec:conclusion}.
    
    \section{Susceptible-infected-recovered-dead (SIRD) model}
    \label{sec:sird_model}

    The susceptible-infected-recovered (SIR) model is a minimal model of epidemic spreading and the most extensively studied model both in complex networks~\cite{Ji2014,moreno2002,Pastor-Satorras2001,Scarpino2019} and in the metapopulation model~\cite{Watts2005,Colizza2007,Colizza2008,Lloyd2004,Masuda2010}, together with its variants~\cite{Choi2017,Cai2015,Chen2013,Cho2009,Bianconi2020,Choi2020,Scarselli2021,Li2022,St-Onge2022,Liu2022}.
    In the SIR model, each individual is in either the susceptible (S), infected (I), or recovered (R) state.
    A susceptible individual can turn into an infected state if it comes into contact with an infected individual.
    If a susceptible individual and an infected individual are in contact, the susceptible individual is turned into the I state at rate $\eta$ (it turns with probability $\eta\Delta t$ in an infinitesimal time step $\Delta t$).
    If the S individual is in contact with $n$ infected individuals, the rate becomes $n\eta$.
    Infected individuals eventually turn into the recovered state at a constant rate $\mu$.
    A recovered individual obtains immunity and does not turn into the infected state again.
    In actual epidemic diseases, there is a probability of reinfection whose effects on the results of this research are discussed in Sec.~\ref{sec:complex_stage}.
    
    Vaccination strategy is one of the core topics in mathematical epidemiology; therefore, considerable research has been devoted to the subject ~\cite{Pastor-Satorras2002,Hebert-Dufresne2013,Cohen2003,Ghalmane2019,Osat2017,Schneider2011,Costa2020,Yan2015,Clusella2016,Masuda2009,VanMieghem2011,Matamalas2018a,Dong2020,Chen2008,Madar2004}.
    The objective of vaccination strategies in the SIR model is to minimize the total number of individuals affected by the disease, which can be measured by the number of recovered individuals when the infection vanishes, with limited vaccination resources.
    However, one of the most important objectives of vaccinations in the real world is to minimize the total number of deaths caused by a disease.
    Because recovery and death are not distinguished in the SIR model, it cannot be used to study vaccination strategies related to such a purpose.    
    At this point, we employ the SIRD model, which is a minimal model that distinguishes recovery and mortality~\cite{Wang2019,Sen2021,Yuan2020,Anastassopoulou2020}.

    In the SIRD model, similar to the SIR model, each individual is in either a susceptible (S), infected (I), recovered (R), or dead (D) state.
    The contagion occurs identically as in the SIR model.
    Any individual from subpopulation $\alpha$ (such as an age group) that is in the I state turns into the R state at rate $(1-\kappa_\alpha)\mu$ or into the D state at rate $\kappa_\alpha\mu$.
    The rate equation for the SIRD model is, therefore,
    \begin{align}
        \mathrm{S}+\mathrm{I} \overset{\eta}{\rightarrow} \mathrm{I}+\mathrm{I} \label{eq:rate_equation_1}\,,\\
        \mathrm{I} \overset{(1-\kappa_\alpha)\mu}{\rightarrow} \mathrm{R} \,,\\
        \mathrm{I} \overset{\kappa_\alpha\mu}{\rightarrow} \mathrm{D} \label{eq:rate_equation_3}\,,
    \end{align}
    where $\eta$ is the contagion rate, $\mu$ is the recovery rate, and $\kappa_\alpha$ is the IFR of subpopulation $\alpha$.
    If an individual from subpopulation $\alpha$ is infected, the individual turns into R state or D state with probability ratio $(1-\kappa_\alpha):\kappa_\alpha$.
    We assumed that the three processes (contagion, recovery, and death) occur independently at constant rates.
    This assumption reasonably describes the pathology of each individual; however, complex social interventions such as quarantine and social distancing that depend on the number of epidemic cases and mortality can complicate the process.

    \section{Results}

    \subsection{Fatality- and contact-based strategies}
    \label{sec:strategies}

	\begin{figure*}
	    \centering
	    \includegraphics{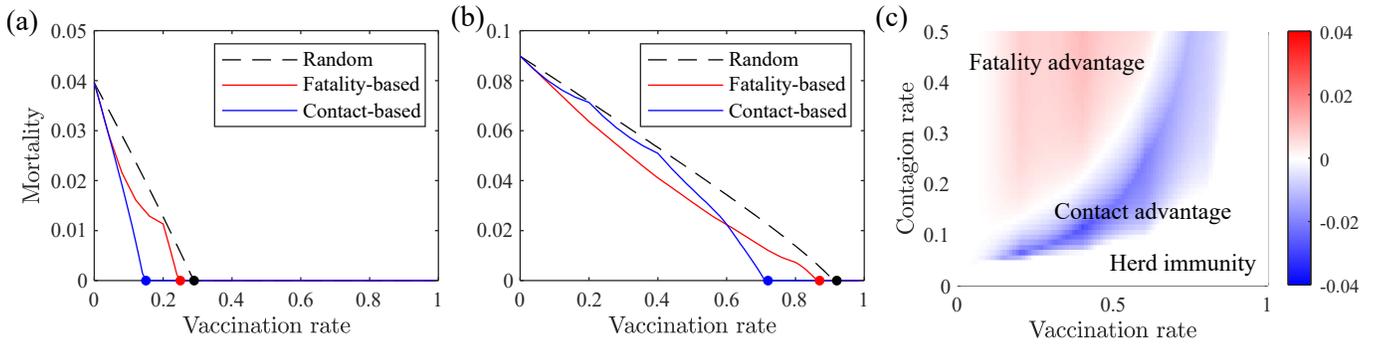}
	    \caption{(a--b) The mortality rate as a result of various vaccination strategies.
	    The contagion rate is (a) $\eta=0.05$ and (b) $\eta=0.4$, while the recovery rate is normalized $\mu=1$.
	    When the fraction of the population affected by the epidemic (i.e., the fraction of R and D state at the end of the dynamic) is less than $10^{-4}$, we assume that herd immunity is achieved, and the epidemic is eliminated by vaccination.
	    The point at which herd immunity is achieved is depicted as a dot.
	    When the contagion rate is low, the contact-based strategy is more effective regardless of the vaccination rate.
	    When the contagion rate is high, the fatality-based strategy is more effective at a low vaccination rate; however, the contact-based strategy outperforms the fatality-based strategy when the vaccine supply is sufficiently high, achieving herd immunity at a lower vaccination rate.
        (c) The difference between the mortality rates resulting from fatality-based and contact-based strategies.
        The fatality-based strategy reduces more deaths compared to the contact-based strategy when the contagion rate is high and the vaccination rate is low.
        However, as the vaccination rate becomes higher, the contact-based strategy outperforms the fatality-based strategy.
        }
	    \label{fig:mortality_synthetic}
	\end{figure*}
    
    A metapopulation model consists of interacting subpopulations, which are often but not necessarily, spatially structured.
    The subpopulations are assumed to be well-mixed.
    For epidemic studies using a metapopulation model, the density of epidemic states in each subpopulation is tracked instead of tracking the epidemic states of each individual.
    The density of states evolves due to the interactions among subpopulations and interactions that occur within the same subpopulation.
    Because metapopulation models have lower dimensions compared to networks, they allow more exhaustive studies on the spread of epidemic diseases.
    The epidemic equation for the SIRD model in the metapopulation model is
    \begin{align}
        \frac{\partial}{\partial t}\rho_{\alpha}^\mathrm{I}(t) &= \eta \left(\rho_\alpha^\mathrm{S}(t) - v_\alpha \right) \sum_\beta M_{\alpha\beta} \rho_\beta^\mathrm{I}(t) - \mu \rho_\alpha^\mathrm{I}(t) \label{eq:metapopulation_homogeneous_1}\,,\\
        \frac{\partial}{\partial t} \rho_\alpha^\mathrm{R}(t) &= \left(1-\kappa_\alpha\right)\mu \rho_\alpha^\mathrm{I}(t) \,,\\
        \frac{\partial}{\partial t} \rho_\alpha^\mathrm{D}(t) &= \kappa_\alpha\mu \rho_\alpha^\mathrm{I}(t) \label{eq:metapopulation_homogeneous_3}\,,
    \end{align}
    where $\rho_\alpha^\mathrm{S}$, $\rho_\alpha^\mathrm{I}$, $\rho_\alpha^\mathrm{R}$, and $\rho_\alpha^\mathrm{D}$ are the probabilities that an individual in group $\alpha$ is in the S, I, R, and D state, respectively; $v_\alpha$ is the fraction of vaccinated individuals in subpopulation $\alpha$; and $M_{\alpha\beta}$ is the contact matrix, which is defined as the average contacts that an individual in group $\alpha$ has with the individuals in group $\beta$.

    Initially, an infinitesimal fraction, $n_0=10^{-8}$ of each group $\alpha$ of the population, is in the I state, and all the rest of the population, $1-n_0$, is in the S state.
    As long as $n_0$ is small enough, the value of $n_0$ and how these initially infected individuals are distributed among the subpopulations do not affect the final states $\rho^\mathrm{R}$ and $\rho^\mathrm{D}$.
    The differential equations are then solved by the fourth order Runge-Kutta method~\cite{Runge1895,Kutta1901} until the total fraction of infected individuals, $\rho^\mathrm{I} = \sum_{\alpha} P_\alpha \rho_\alpha^\mathrm{I}$, becomes less than a certain threshold, $10^{-12}$, and the epidemic process ends ($P_\alpha$ is the fraction of individuals in subpopulation $\alpha$.).
    We then calculate the total fraction of the deceased population $\rho^\mathrm{D} = \sum_\alpha P_\alpha \rho_\alpha^\mathrm{D}$.
        
    We constructed a metapopulation model with heterogeneous contact and fatality rates.
    The population has fatality rates $\kappa_i=$5\%, 7.5\%, 10\%, 12.5\%, and 15\% and relative contact rates $c_j=$0.5, 0.75, 1, 1.25, and 1.5.
    The population is equally divided into 25 subpopulations according to the five fatalities and five contact rates ($5\times 5=25$).
    The contact rate between groups $(i,j)$ and $(i^\prime,j^\prime)$ is $M_{iji^\prime j^\prime}=c_jc_j^\prime$.
    
    We investigated the effectiveness of random, fatality-based, and contact-based strategies for various levels of vaccine supply.
    In the random strategy, the vaccine is randomly distributed and each subpopulation is uniformly vaccinated.
    In the fatality-based strategy, the subpopulations are vaccinated in descending order of fatality rates, and if two subpopulations have identical fatality rates, the one with a higher contact rate is vaccinated.
    In the contact-based method, an infinitesimal amount of vaccine is iteratively given to the age group with the highest contact rate with unvaccinated individuals until the total amount of vaccine is distributed.
    The contact rate of age group $\alpha$ with unvaccinated individuals is 
    \begin{equation}
        \sum_\beta M_{\alpha\beta}\left( 1-v_\beta \right)\,,
    \end{equation}
    and the value is recalculated at each iteration.
    This differs from the contact rate of age group $\alpha$ with any individual in the population, which is $\sum_\beta M_{\alpha\beta}$.

    The mortality rate of the population when each vaccination strategy is employed is illustrated in Figs.~\ref{fig:mortality_synthetic}(a) and (b).
    When the contagion rate is low, the contact-based strategy results in a lower mortality rate than the fatality-based strategy regardless of the vaccination rate; however, for a high contagion rate, there is a crossover between the strategies.
    The fatality-based strategy more effectively reduces mortality when the vaccination rate is low, but the contact-based strategy outperforms the fatality-based strategy when the vaccination rate is high.
    If the vaccination rate is sufficiently high, herd immunity is achieved regardless of the choice of the vaccination strategy (fatality-based, contact-based, random, etc).
    The difference between the mortality rates when fatality- and contact-based strategies are employed is depicted in Fig.~\ref{fig:mortality_synthetic}(c).
    The fatality-based strategy is effective when the contagion rate is high and the vaccine supply is low.
    
    \begin{figure*}
	    \includegraphics{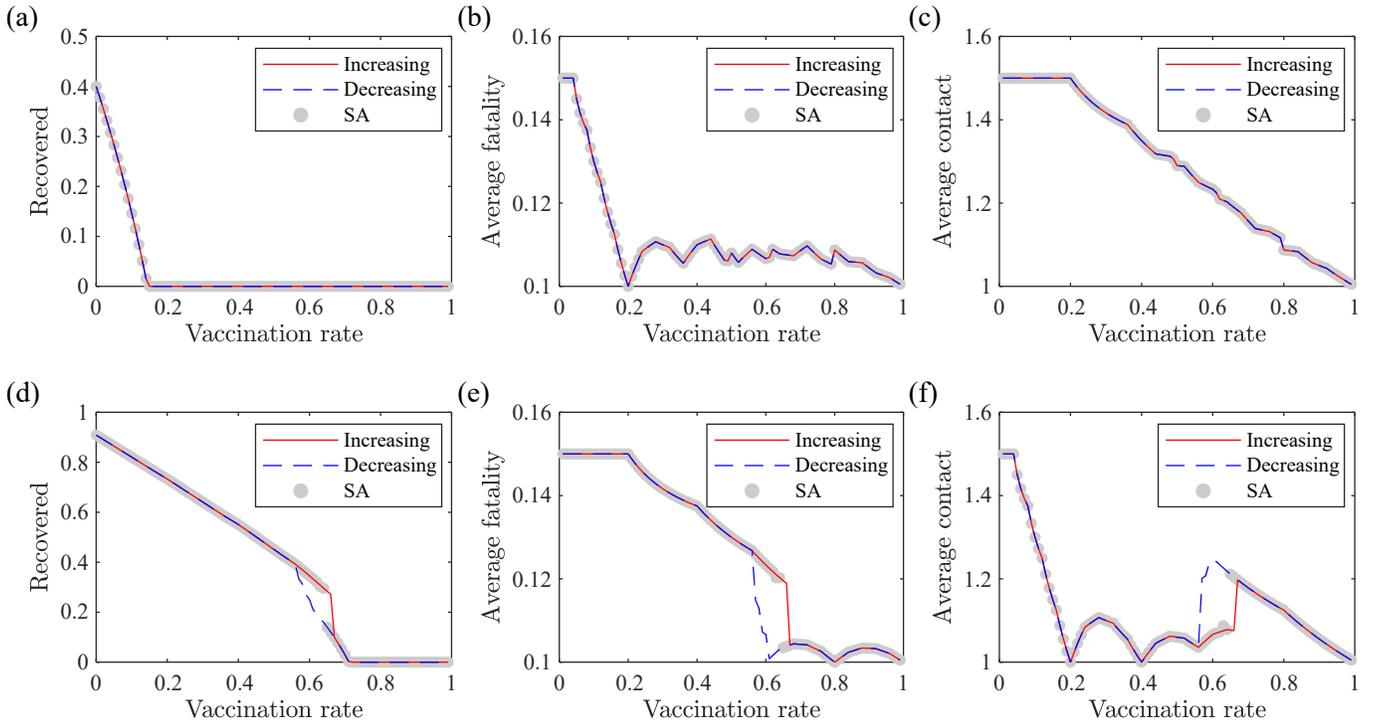}
        \caption{(a, d) Fraction of recovered population, (b, e) average fatality of the vaccinated population, and (c, f) average contact rate of the vaccinated population of the synthetic metapopulation model with heterogeneous fatality and contact rates.
        The recovery rate is normalized to $\mu=1$, and contagion rates are (a--c) $\eta=0.05$ and (d--f) $\eta=0.4$.
        The increasing (decreasing) curve, where the locally optimal vaccination strategies are found by iteratively increasing (decreasing) the vaccination rate, is depicted as solid red (dashed blue) lines.
        The globally optimal strategies are found by simulated annealing (SA).
        There is no abrupt transition nor separation of the curves for a low contagion rate, $\eta=0.05$, and the optimal vaccination strategy prefers to vaccinate individuals with high contact rates regardless of the vaccination rate.
        For a large contagion rate, $\eta=0.4$, there is an abrupt transition in the globally optimal strategy.
        The separation of the increasing and decreasing curve indicates the path-dependency of the vaccination strategy which is analogous to hysteresis.
        For a small (0--0.56) or large (0.67--1) vaccination rate, the increasing and decreasing curves coincide; however, near the transition point, the increasing curve tends to vaccinate individuals with high fatality rates (high-fatality strategy) and the decreasing curve vaccinates individuals with high contact rates (high-contact strategy).
        The High-fatality strategy results in a higher number of recovered population than the high-contact strategy because the high-fatality strategy aims to protect high-risk groups, while the high-contact strategy aims to contain the infection.}
        \label{fig:hysteresis_synthetic}
    \end{figure*}
    
    \begin{figure}
        \centering
	    \includegraphics{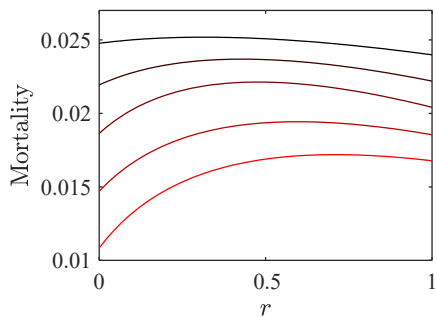}
        \caption{The mortality rate of a mixed strategy that combines the high-fatality and high-contact strategies.
        The contagion rate is $\eta=0.4$, and vaccine supply is, from top to bottom, 58\%, 60\%, 62\%, 64\%, and 66\%. 
        There is a barrier of mortality rate between the high-fatality and high-contact strategies, which is the cause of the path-dependency and discontinuous transition of the vaccination strategy.}
        The mortality rate of the high-contact strategy drops faster than the high-fatality strategy as the level of vaccine supply increases, causing the crossover between the two strategies.
        \label{fig:barrier_synthetic}
    \end{figure}

    \subsection{Transition and path-dependency of the optimal vaccination strategy}
    \label{sec:phase_transition}

    In this section, we further investigate the vaccination rate dependency of the vaccination strategy and demonstrate that the optimal vaccination strategy undergoes a discontinuous transition.
    To find the globally optimal vaccination strategy, we implement a modified version of the simulated annealing technique.
    The simulated annealing is a probabilistic optimization algorithm inspired by spin glass~\cite{Kirkpatrick1987}.
    First, we start with a random vaccination strategy with a given amount of vaccine supply.
    We set this strategy as a provisional solution.
    We then calculate the mortality rate of a trial strategy, which is perturbed from the provisional solution by a small amount while keeping the vaccine supply of the total population constant.
    If the mortality rate of the trial strategy is smaller than that of the provisional solution, we replace the provisional solution with the trial strategy.
    Otherwise, we replace the provisional solution with the trial strategy with probability $\exp{\left(-1/T\right)}$, where $T$ is the temperature of the algorithm.
    In the beginning, the temperature is set at $T=2$.
    We iterate this process $n_{\mathrm{iter}}=10^6$ times, while the temperature is dropped by a factor of $f_{\mathrm{iter}}=1-2\times10^{-5}$ at each step.
    The resulting provisional solution is the optimal vaccination strategy, given that $n_{\mathrm{iter}}$ is sufficiently large and $f_{\mathrm{iter}}$ is sufficiently close to one.    
    To find locally optimal solutions, we use the zero-temperature simulated annealing, which is analogous to the gradient-descent method.
    We perturb the provisional solution by decreasing the vaccination rate of group $\alpha$ by $\delta v/P\left(\alpha\right)$ and increasing the vaccination rate of group $\beta$ by $\delta v/P\left(\beta\right)$, where $\delta v$ is a small number, and $P\left(\alpha\right)$ is the fraction of the group $\alpha$ in the population.
    This way, the total vaccination rate of the entire population remains constant.
    Among perturbed solutions, if any solution results in a smaller mortality rate, we replace the provisional solution with the perturbed solution that results in the smallest mortality rate.
    Otherwise (i.e., if all the perturbed solutions result in larger mortality rates than the provisional solution), we have achieved a locally optimal solution; hence, we terminate the process.    
    
    To investigate the path-dependency of the optimal vaccination strategy, we iteratively increased and decreased the vaccination rate by a small amount, while constantly calculating the locally optimal vaccination strategy in the vicinity.
    To obtain the increasing curve, we first set the vaccination rate to $\Delta v=0.01$ and find the optimal vaccination strategy $v^{(I)}_\beta(\Delta v)$.
    We then increase the vaccination rate by $\Delta v$ and find the locally optimal vaccination strategy $v^{(I)}_\beta(2\Delta v)$ near the optimal strategy from the previous step.
    We repeat this process until the vaccination rate reaches one.
    To obtain the decreasing curve, we start from a vaccination rate of $1-\Delta v$ and repeat the process.

    The results are illustrated in Fig.~\ref{fig:hysteresis_synthetic}.
    Fraction of the recovered population $\rho^\mathrm{R}$, average fatality, and contact rates of the vaccinated individuals are depicted as characteristics of vaccination strategies.
    These values are analogous to the order parameters of the phase transition in thermal systems and the vaccination rate is the control parameter.
    The order parameters of the two local mortality minima are depicted in the curves similarly to the magnetization of the two free energy minima is depicted in the hysteresis curve of the magnetic systems.
    The global mortality minimum corresponds to the global free energy minimum where the system lies in the Boltzmann distribution.
    For a low contagion rate, there is no abrupt transition of the optimal vaccination strategy.
    For a high contagion rate, an abrupt transition of the globally optimal vaccination strategy, which is obtained by simulated annealing, discontinuously changes.
    Moreover, there is a path-dependency in the vaccination strategy.
    When locally optimal vaccination strategies are found by slowly increasing the vaccination rate from zero, the strategies vaccinate individuals with high fatality rates in the middle region (high-fatality strategy).
    If the strategies are found by slowly decreasing the vaccination rate from one, they primarily vaccinate individuals with high contact rates (high-contact strategy).
    A high-fatality strategy results in a higher fraction of recovered individuals than the high-contact strategy, even though the strategies' mortality rates are similar or the same in the vicinity of the transition point.
    
    The path-dependency of this transition implies that a moderate strategy that combines the high-fatality and high-contact strategy can be less effective than either strategy.
    The vaccination rate of the moderate strategy is $v_\alpha^{\mathrm{mod}} = r v_\alpha^{\mathrm{f}} + (1-r) v_\alpha^{\mathrm{c}}$, where $v_\alpha^{\mathrm{f}}$ and $v_\alpha^{\mathrm{c}}$ are the vaccination rates of subpopulation $\alpha$ in the high-fatality and high-contact strategy, respectively.
    The performance of the moderate strategy for various levels of vaccine supply is depicted in Fig.~\ref{fig:barrier_synthetic}.
    There is a barrier of mortality rate between the high-fatality and high-contact strategies, and the moderate strategy is never more effective than both of the strategies, and in some regions, it is less effective than either of the two strategies.
    Hence, it is inadvisable to mix the two strategies or change from one to the other in the middle.
    The mortality rate of the high-contact strategy ($r=0$) decreases faster than the high-fatality strategy ($r=1$), which results in an abrupt transition of the optimal vaccination strategy from a high-fatality to a high-contact strategy.

    \subsection{Real-world epidemic diseases}
    \label{sec:real_world}

    \begin{figure}
        \centering
	    \includegraphics[width=\columnwidth]{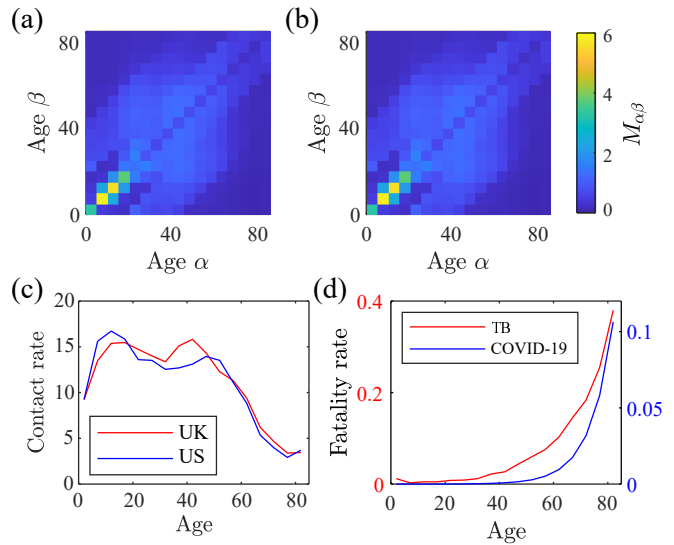}
		\caption{
		The age contact matrix of the (a) United Kingdom and (b) United States, and (c) contact rate of each age group.
		The population is divided into 17 groups: aged 0--4, 5--9, \dots, 75--79, and above 80.
		The interaction strength between groups of similar age is disproportionately higher, and the groups of ages 10--29 show the highest number of contacts.
		(d) The fatality rates of TB in the United Kingdom (UK) and COVID-19 in the United States (US).
		The fatality rate is highly heterogeneous and monotonically increases with age, with the sole exception of children below age five in TB.
		Therefore, the senior population is primarily vaccinated by the fatality-based strategy, and individuals of age 10--29 are primarily vaccinated by the contact-based strategy.
		}
		\label{fig:dataset}
	\end{figure}
    
    \begin{figure*}
        \centering
	    \includegraphics{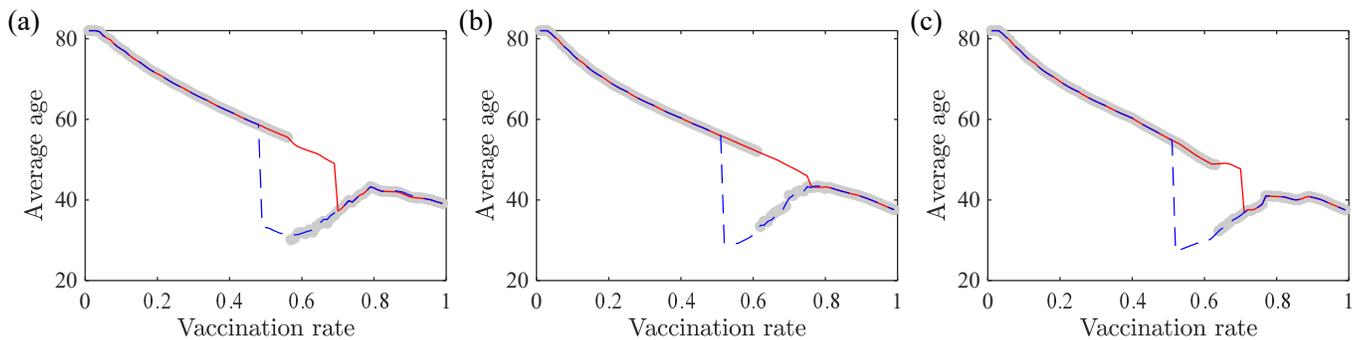}
        \caption{Average age of vaccinated individuals for (a) TB, (b) COVID-19, and (c) COVID-19 with reinfection and finite vaccine efficacy.
        The contagion rates are $\eta=0.25$, and the recovery rate is normalized to $\mu=1$.
        The total amount of vaccine is increased (solid red line) and decreased (dashed blue line) while keeping the vaccination strategy at its local optimum.
        For a small vaccination rate, the increasing curve and decreasing curve coincide; however, for a sufficiently large vaccination rate, the two curves show a largely unequal average age of the vaccinated population.
        The increasing curve has a higher average age.
        The globally optimal strategy, which is calculated by the simulated annealing, undergoes an abrupt transition from the high-fatality to the high-contact strategies.
        The separation of the increasing and decreasing curve indicates the path-dependency of the vaccination strategy which is analogous to hysteresis.}
        \label{fig:hysteresis_real}
    \end{figure*}
    
    \begin{figure}
        \centering
	    \includegraphics{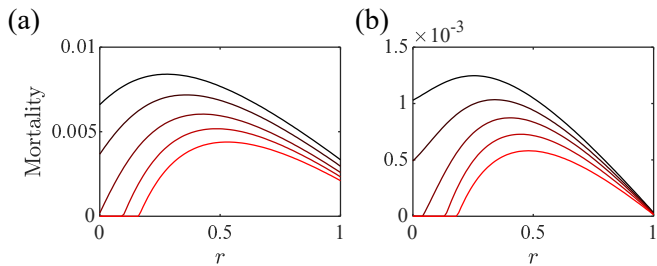}
        \caption{The mortality rate of the mixed strategy of high-fatality and high-contact strategies for (a) TB and (b) COVID-19.
        The contagion rate is $\eta=0.25$, and the vaccine supply is, from top to bottom, 52\%, 54\%, 56\%, 58\%, and 60\%. 
        There is a barrier of mortality rate between the high-fatality and high-contact strategy, which is the cause of the discontinuous transition and path-dependency of the optimal vaccination strategy.
        The mortality rate of the high-contact strategy drops faster than the high-fatality strategy as the level of vaccine supply increases, and herd immunity is achieved at a lower vaccine supply.
        }
        \label{fig:barrier_real}
    \end{figure}
    
    In this section, we show that the discontinuous transition and path-dependency of the optimal vaccination strategy illustrated in the previous section occur in the vaccination of actual epidemic diseases, such as tuberculosis (TB) and COVID-19.
    Contact data between each age group in the various countries have been studied utilizing surveys~\cite{Mistry2021}.
    To model TB, we employed a contact matrix calculated from the UK data along with the incidence risk ratio of TB in the UK~\cite{Pedrazzoli2019}.
    The population is divided into 17 groups: aged 0--4, 5--10, ..., 75--79, and above 80.
    For COVID-19, we use the contact matrix of the US and the age-dependent IFR obtained from a meta-analysis of medical literature~\cite{Levin2020}. The latter is calculated as
    \begin{equation}
        \log_{10} {\rm IFR} = (-3.27 \pm 0.07) + (0.0524 \pm 0.0013) \, {\rm age} \,.
    \end{equation}
    The contact matrices and the fatality rates of the diseases are illustrated in Fig.~\ref{fig:dataset}.
    Contact within a similar age group is disproportionately intense (Figs.~\ref{fig:dataset}(a, b)), and contacts between teenagers exhibit the highest strength.
    Fatality rates of the diseases monotonically increase with age, except for children below age five for TB.
    As a result, the fatality-based strategy primarily vaccinates the senior population, while the contact-based strategy vaccinates the teenagers first.
    
    The average age of the vaccinated individuals are presented for TB (Fig.~\ref{fig:hysteresis_real}(a)) and COVID-19 (Fig.~\ref{fig:hysteresis_real}(b)).
    Both figures exhibit the discontinuous transition and path-dependency of the optimal vaccination strategy.
    The increasing curve vaccinates the senior population more than the decreasing curve, which corresponds to the phenomenon depicted in Fig.~\ref{fig:hysteresis_synthetic}(e), where the increasing curve has a greater preference to vaccinate individuals from groups with high fatality rates than the decreasing curve.
    Also, there are barriers of mortality rate between the high-fatality and high-contact strategies (Fig.~\ref{fig:barrier_real}), suggesting that mixing the two strategies is ineffective.
    As the vaccination rate increases, the mortality associated with the high-contact strategy decreases faster than that of the high-fatality strategy to achieve herd immunity at a lower vaccine supply.
    
    \subsection{Complex epidemic stages, vaccine breakthrough infection, and reinfection}
    \label{sec:complex_stage}
    
    The previous results are obtained in simplified models.
    In this section, we demonstrate that the more complicated behaviors of the actual diseases do not significantly alter the findings of this research. 
    First, in the real world, actual infectious diseases progress in a series of epidemic stages, such as the incubation period, prodromal period, and acute period.
    Each of these stages has a distinct rate of spreading the disease.
    These stages have complicated effects on the temporal dynamics of epidemics, but in this study, only the fraction of population in each epidemic state (R and D) at the end of the epidemic is relevant.
    In this sense, the complex stages of a disease can be reduced to a simplified model.
    For instance, suppose there multiple infectious stages $\mathrm{I}_k$ ($k=1,\cdots,K$) of a disease, each with contagion rate $\eta_k$ and progression rate $\mu_k$ (i.e., $\mathrm{S}+\mathrm{I_k} \rightarrow \mathrm{I}_1+\mathrm{I}_k$ occurs with rate $\eta_k$, $\mathrm{I}_k \rightarrow \mathrm{I}_{k+1}$ occurs with rate $\mu_k$, and $\mathrm{I}_K \rightarrow \mathrm{R}$ occurs with rate $\mu_K$).
    The total recovered population at the end of the epidemic disease is then identical to that of the SIR model with contagion rate $\eta^{*}=\sum_k \eta_k/\mu_k$ and $\mu=1$.
    To model an incubation stage, we can set $\eta_I=0$ and $\mu_I=1/\tau_I$, where $\tau_I$ is the incubation period of the disease.
    
    Also, we assumed that vaccinated individuals never become infected even in contact with infected individuals.
    However, people who are vaccinated still can get infected by COVID-19.
    An infection of a vaccinated individual is referred to as a vaccine breakthrough infection.
    To include vaccine breakthrough infection, we can suppose a vaccine efficacy of $\theta<1$, and the vaccinated individuals turn into the infected state at a rate of $(1-\theta)\eta$ instead of $\eta$.
    Additionally, even when an infected individual recovers and obtains immunity to the disease, there is a small probability that the individual can be infected by the disease again.
    Such reinfection can be modeled as some individuals losing immunity~\cite{Okuwa2019,Saif2019,Salman2021}.
    Hence, individuals in state R turn into S state at rate $\nu$.
    The typical time for an individual to lose immunity is $1/\nu$.
    Individuals in the D state remain in the D state.
    
    We included vaccine breakthrough infections and reinfections to COVID-19 and illustrated the average age of the vaccinated population in Fig.~\ref{fig:hysteresis_real}(c).
    The vaccine efficacy is $\theta=0.9$, and the rate of immunity loss is $\nu=0.05$.
    This means that typically the immunity of a vaccine is lost over a duration 20 times the average recovery time of the disease ($\sim 200$ days).
    The discontinuous transition of the optimal strategy and path-dependency still manifests themselves in the model with these modifications.
    
    \section{Conclusion}
    \label{sec:conclusion}

    In summary, we employed the SIRD model to investigate the effectiveness of vaccination strategies to minimize the mortality rate in a population with heterogeneous fatality rates.
    We constructed a synthetic metapopulation model with heterogeneous fatality and contact rates to investigate how the effectiveness of vaccination strategies relates to the amount of vaccine available.
    Vaccinating individuals with high fatality rates is effective when the contagion rate is high and the vaccine supply is low.
    We found the discontinuous transition and path-dependency, which is analogous to hysteresis in statistical physics, of the optimal vaccination strategy.
    The path-dependency of the vaccination strategy implies that combining high-fatality and high-contact strategies is ineffective in reducing the mortality rate of the epidemic disease.
    We also demonstrated that such phenomena occur in real-world epidemic diseases, such as TB and COVID-19.
    These conclusions are valid even when complex stages of a disease, vaccine breakthrough infection, and reinfection are considered.
    
    In conclusion, the effectiveness of vaccination strategies is closely related to the amount of vaccine available.
    Hence, the quantity of vaccine supply should be estimated before the design of the vaccination strategies.
    Precise estimation of the contact matrix, basic reproduction number, and the IFR of the population is also important.
    In the survey data used in this paper, all types of contacts were treated equally.
    However, the contagion rate of disease among individuals who live in the same house, work in the same place, or shop in the same grocery store should differ from each other.
    If more accurate contagion tree data of the disease are collected and implemented, the relative strength of such interactions can be taken into account.
    Although the effectiveness of the strategies at specific vaccination rates will be modified if the precision of the dataset is improved, because the discontinuous transition and path-dependency of the optimal vaccination strategies occur in various epidemic models with a wide range of parameters, the conclusions of this research should still be valid.    

    \begin{acknowledgments}
        This research was supported by the NRF, Grant No.~NRF-2014R1A3A2069005.
    \end{acknowledgments}
	
    \section*{Data Availability Statement}

    The data that support the findings of this study are openly available in \url{https://github.com/mobs-lab/mixing-patterns}~\cite{Mistry2021} (contact matrices), Ref.~\cite{Pedrazzoli2019} (age-specific fatality of TB), and Ref.~\cite{Levin2020} (age-specific fatality of COVID-19).

\end{document}